\documentclass[runningheads]{llncs}

\usepackage[utf8]{inputenc}
\usepackage{graphicx}
\usepackage{amsmath}
\usepackage{amsfonts}
\usepackage{amssymb}
\usepackage{booktabs}
\usepackage{times}
\usepackage{subcaption}
\usepackage[colorlinks]{hyperref}
\usepackage{longtable}
\usepackage{multirow}
\usepackage{parskip}
\usepackage{paralist}
\usepackage{kantlipsum}

\newcommand{\ballots}{\mathcal{B}}

\usepackage{todonotes}

\newcommand{\ignore}[1]{}

    {\begin{center}\begin{minipage}{\textwidth}\begin{tabbing}}%
    {\end{tabbing}\end{minipage}\end{center}}

\title{Random errors are not necessarily politically neutral\thanks{A later version of this work appears in E-Vote-ID 2020.}}

\author{
Michelle Blom    \inst{1}   \orcidID{0000-0002-0459-9917}  \and
Andrew Conway               \orcidID{0000-0001-6277-2442}  \and
Peter J. Stuckey \inst{2}   \orcidID{0000-0003-2186-0459}  \and
Vanessa Teague   \inst{3,4} \orcidID{0000-0003-2648-2565}  \and
Damjan Vukcevic  \inst{5,6} \orcidID{0000-0001-7780-9586}}

\authorrunning{Blom et al.}
\institute{
School of Computing and Information Systems, University of Melbourne,
Parkville, Australia
\and
Faculty of Information Technology, Monash University, Clayton, Australia
\and
Thinking Cybersecurity Pty.\ Ltd.
\and
College of Engineering and Computer Science, Australian National University
\and
School of Mathematics and Statistics,
University of Melbourne, Parkville, Australia
\and
Melbourne Integrative Genomics,
University of Melbourne, Parkville, Australia}

\begin{document}

\maketitle

\begin{abstract}
Errors are inevitable in the implementation of any complex process.  Here we
examine the effect of random errors on Single Transferable Vote (STV)
elections, a common approach to deciding multi-seat elections. It is usually
expected that random errors should have nearly equal effects on all candidates,
and thus be fair.  We find to the contrary that random errors can introduce
systematic bias into election results. This is because, even if the errors are
random, votes for different candidates occur in different patterns that are
affected differently by random errors.  In the STV context, the most important
effect of random errors is to invalidate the ballot.  This removes far more
votes for those candidates whose supporters tend to list a lot of preferences,
because their ballots are much more likely to be invalidated by random error.
Different validity rules for different voting styles mean that errors are much
more likely to penalise some types of votes than others.  For close elections
this systematic bias can change the result of the election. 
\end{abstract}

% -----------------------------------------------------------------------------

\section{Introduction}

We investigate the effects of random errors on election outcomes, in the
context of preferential elections counted using the Single Transferable Vote
(STV).  It is often assumed that random errors (whether from human or manual
counting) are unimportant because they are likely to have nearly equal effects
on all candidates.  In this paper we show that this is not the case, using
simulated random errors introduced into real STV voting data. In some cases,
this introduces a systematic bias against some candidates.

Random errors have a non-random effect because real votes are not random.
Voters not only express different preferences, but express them in a different
way, according to whom they choose to support.

In STV, some candidates are elected mainly on the strength of their party
listing; others rely on gathering preference flows from other parties, or on
their individual popularity relative to their party's other candidates.  So
when we look at the votes that contributed to the election of different
candidates, we find that the types of votes chosen by their supporters may be
very different.  Hence a random error that affects different types of votes
differently introduces a systemic change in the election result.

One obvious kind of error is to misrecord a number.  Usually, this either
invalidates the ballot completely, or invalidates preferences below the error.
The more preferences there are on a ballot, the more likely that at least one
of them is misrecorded.  So as a general rule, candidates that are more
dependent on later preferences or long preference lists are more severely
disadvantaged by random errors. 

Although these results are significant, and need to be taken into account for
close contests, we find that reasonable error rates produce changes in only
very few elections, which (so far) correspond only to those that are obviously
very close.  It is possible for STV elections to have hidden small margins, but
this seems to be uncommon---in almost all the elections we simulated, no
plausible error rate produced a change in outcome.  Typical random error rates
will affect election results when the election is close, but are not expected
to do so when the election is not close.

We do not consider the errors necessary to alter the election result in a
targeted way by altering specific carefully chosen votes---they would obviously
be much smaller.  Hence the results of this paper apply to random errors, but
not deliberate electoral fraud.

The remainder of the paper is organized as follows.  In the next section we
explain STV elections, in particular in the case of Australian Senate
elections, and discuss how the votes are digitised and counted.  In
\autoref{sec:design} we describe our experiment design and introduce the three
error models we explore.  In \autoref{sec:rate} we provide a number of
different approaches to estimate the likely error rate that occurs for
Australian Senate elections.  In \autoref{sec:results} we examine the result of
applying simulated errors to Australian Senate elections and discuss how these
errors can change the result of the election.  Finally in \autoref{sec:conc} we
conclude.

% -----------------------------------------------------------------------------

\section{Background on STV counting}

\subsection{The Single Transferable Vote (STV) counting algorithm}
\label{sec:STVBackground}

STV is a multi-winner preferential voting system.  Candidates compete for $s$
available seats. A candidate is awarded a seat when their tally reaches or
exceeds the quota, $Q$, defined as a function of the number of ballots cast in
the election, $|\ballots|$, and the number of seats, $s$. One popular
definition is the Droop quota,
\begin{equation*}
    Q = \left\lfloor \frac{|\ballots|}{s + 1} \right\rfloor + 1.
\end{equation*}

\begin{figure}[t]
    \centering
    \begin{subfigure}{\linewidth}
    \centering
    \includegraphics[height=4.5cm]{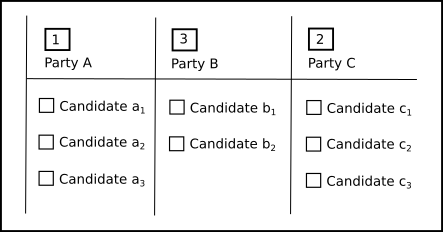}
    \caption{}
    \label{fig:BallotATL}
    \end{subfigure}
    \begin{subfigure}{\linewidth}
    \centering
    \includegraphics[height=4.5cm]{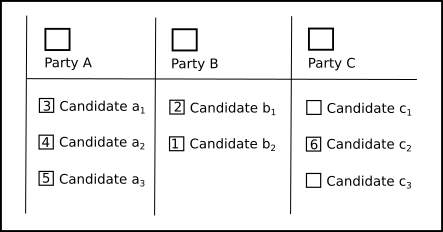}
    \caption{}
    \label{fig:BallotBTL}
    \end{subfigure}
    \caption{\textbf{An example of two simple ballots for a 3-party STV
    election.} In (a), the voter has chosen to vote above the line, and in (b)
they have voted below the line.}
    \label{fig:eg_ballot}
\end{figure}

When a voter casts a ballot in one of these STV elections, they have the option
of voting `above the line' or `below the line'. \autoref{fig:eg_ballot} shows
an example of a ballot for a simple STV election in which candidates from three
parties are competing for $s$ seats. Each party or group of independents
fielding candidates in the election have a box sitting `above the line' (ATL).
A voter may rank these parties and groups by placing a number in their
corresponding box (\autoref{fig:BallotATL}). Alternatively, a voter may rank
individual candidates by placing a number in their box, below the line (BTL)
(\autoref{fig:BallotBTL}). 

Tabulation starts by giving each candidate all the below-the-line ballots in
which they have been ranked first. ATL ballots are awarded to the first
candidate listed under the party that has been ranked first. For example, a
ballot in which Party A has been ranked first sits in the first preference pile
of candidate $a_1$. A BTL ballot in which candidate $b_2$ is ranked first sits
in that candidate's first preference pile. Each ballot is assigned a weight,
starting at 1, that changes as counting proceeds. The tally of a candidate is
the sum of the weights of ballots sitting in their tally pile, possibly with
some rounding.

Counting proceeds by awarding a seat to all candidates whose tallies have
reached or exceeded $Q$. Their \emph{surplus}---their tally after subtracting
$Q$---is distributed to remaining eligible candidates. A candidate is eligible
if they have not been eliminated, and their tally has not reached a quota's
worth of votes. The ballots sitting in an elected candidate's tally pile are
re-weighted so that their combined weight is equal to the candidate's surplus.
These ballots are then given to the next most-preferred eligible candidate on
their ranking. The ATL ballot in \autoref{fig:BallotATL} is given to candidate
$a_2$ if $a_1$ is elected to a seat. If neither $a_2$ or $a_3$ are eligible,
the ballot then moves to candidate $c_1$. The BTL ballot in
\autoref{fig:BallotBTL} is given to candidate $b_1$ if $b_2$ is elected or
eliminated.

If no candidate has reached a quota, the candidate with the smallest tally is
eliminated. The ballots in their tally pile are distributed to the next
most-preferred eligible candidate in their rankings at their current weight.

Counting proceeds in rounds of election and elimination until all $s$ seats are
filled, or until the number of candidates that remain eligible equals the
number of remaining seats. In this setting, each of the remaining candidates is
awarded a seat.

% -----------------------------------------------------------------------------

\subsection{Australian vote digitisation in practice}
\label{subsec:AECProcess}

Australians cast their votes on paper ballots.  The Australian Electoral
Commission (AEC) digitises the preferences in a hybrid manual and automated
process.  Precise details about this process are unavailable, but most ballots
seem to receive both  automated digitisation and secondary human data entry.
(Ballots that are judged blank are not re-examined.) It is possible that manual
data entry is performed on ballot
papers.\footnote{\url{https://www.aec.gov.au/Voting/counting/files/css-integrity.pdf}}
Other pamphlets suggest that only the images, not the paper ballots, are
used.\footnote{\url{https://www.aec.gov.au/Voting/counting/files/senate-count.pdf}}

An automated system then checks, for each ballot, whether the automated
digitisation matches the human interpretation.  Obviously this does not defend
against software errors or deliberate manipulation, particularly downstream of
the process, but it probably does produce reasonably low random error rates,
assuming that the human errors are not highly correlated with the errors of the
automated system.

Ballots are required to have a minimum number of preferences before they are
considered valid; such ballots are referred to as \emph{formal} ballots.  In
the 2016 and 2019 elections, a BTL formal vote must have every preference from
1 to 6 inclusive present exactly once; an ATL formal vote requires the
preference 1 to be present exactly once and a formal BTL vote not to be
present.
%Voters are told to mark at least 6 preferences
%if voting above the line, and 12 if voting below the line.
According to the information about the digitisation processes mentioned above,
non-blank informal ballots seem to get a second human inspection automatically.

The AEC publishes on their website the complete digitised preferences for all
Senate votes, excluding blanks and votes judged to be informal.

In summary, the published data could differ from the actual ballots for many
reasons:

\begin{itemize}
\item random errors that match in both the automated and human digitisation
    process,
\item random errors that occur in either the automated or human digitisation
    process, and are endorsed rather than corrected by the reconciliation
    process,
\item erroneous exclusion of ballot papers judged to be informal,
\item accidental alterations, duplicates or omissions caused by software bugs,
\item deliberate manipulation by malicious actors, either of the images (before
    digitisation) or of the preference data (from digitisation to
    publication).
\end{itemize}

Our investigation does not apply to the last two kinds of errors, which could
be introduced in a non-random way that worked for or against a particular
candidate.  It does apply to the errors that are random.  In particular, we
show that digitisation errors that randomly  cause some ballots to be judged
informal can impact candidates differently.

% -----------------------------------------------------------------------------

\section{Experimental design}
\label{sec:design}

Our analysis is performed on the AEC's published data for the 2016 and 2019
Australian federal elections for the Senate, i.e.\ the output of the process
described in \autoref{subsec:AECProcess}.
%The analysis for the 2013 and prior elections would be somewhat different as
%the formality requirements on below the line (BTL) votes changed
%significantly, and the meaning of above the line votes (ATL) changed
%massively, from party chosen tickets (2013) to preferential ordering of
%parties (2016).
%%%%
Ideally our analysis would be based upon the actual marks that voters made on
their ballots, or even what they intended to make, and the comparison with the
AEC's output.  However, these data are not available. Instead, we use the AEC's
output as the `actual' ballot data, and add simulated errors.

%The AEC uses slightly different file formats for the two elections, and
%multiple files.

\subsection{Analysis code}

For logistical reasons, and to make it easy for anyone to replicate this
experiment, we extract those preferences that are actually considered valid in
the election. If a number is absent or repeated in the preference marks, then
it and all subsequent preferences are disregarded.  We have made available a
standardised ``.stv'' file format based on the data published by the
AEC\footnote{See the downloads section for each election at:
\url{https://vote.andrewconway.org}}. This common format does unfortunately
mean that we lose some (invalid) marks that could conceivably have become valid
when we added new random errors, or which could, through errors, invalidate
earlier preferences.

We used the Java pseudo-random number generator \texttt{java.util.Random} to
generate random numbers, and ensured that different executions used different
seeds.  Our code is available for
download\footnote{\url{https://github.com/SiliconEconometrics/PublicService}}.

\subsection{Error models}

We simulate the effect of errors by making random changes to the votes. We are
not certain exactly what ``random'' failures in the scanning process would be,
so we have devised three different models for simulated errors, in increasing
order of complexity and plausibility.  The first models an error where,
somewhere in the list, something goes wrong that invalidates the rest of the
preference list.  The second models an error in which a digit is randomly
misread as another digit, chosen uniformly.  The final model recognises that
some misreadings are much more likely than others---for example, a 3 is more
likely to be confused with an 8 than a 1---so we use a model that includes a
specific error probability for each digit and each potential misreading.

Each model applies to a valid list of preferences and treats either each number
or each digit separately with random errors chosen independently.

\begin{enumerate}
\item For each preference, with probability $\epsilon$, truncate the list at
    that preference.

\item For each digit, with probability $\epsilon$, replace that digit with a
    digit uniformly chosen from $\{0,1,2,3,4,5,6,7,8,9\}$, which may be the
    original digit.

\item Start with a table of pairwise error ratios for digits such as
    \autoref{tab:confus} (that is, the probability that a certain digit is
    mistranscribed into a certain other digit).   For each digit, change it
    into a different digit with the probability given in the table.
\end{enumerate}

Note that in all three models, the probability of at least one error on the
ballot increases with the number of preferences listed on the ballot.  We are
primarily motivated by machine errors, so per-digit or per-number random errors
seem plausible, but it is worth noting that other errors might be important
too, such as models that considered that some voters (those with bad
handwriting) were much more likely to have their vote misinterpreted than
others.

After applying errors, formality rules are checked again, reducing the number
of ballots considered for the election.

% -----------------------------------------------------------------------------

\section{What is a realistic error rate?}
\label{sec:rate}

As far as we know, there are no publicly available results from any rigorous
estimate of Senate scanning random errors in Australia.  However, there are
several independent estimates, which give us a per-digit error rate ranging
from 0.01\% to 0.38\%.
We define an error to be a discrepancy between the paper ballot and the
electronic preference list output at the end of the process.

\subsection{Using data from the Australian Electoral Commission}

As far as we know, the AEC does not conduct, nor allow anyone else to conduct,
a large random sample of Senate ballots for comparison between electronic and
paper records.  However, an Australian National Audit Office
report\footnote{\url{https://www.anao.gov.au/work/performance-audit/aec-procurement-services-conduct-2016-federal-election}}
describes a process for gaining an estimate from a small sample. This process
was conducted by AEC officials.

\begin{itemize}
    \item A batch of 50 ballot papers was randomly selected and then six ballot
        papers from that batch were reviewed;
    \item Compliance inspectors recorded the first six preferences from the
        physical ballot paper on a checklist;
    \item Verification officers compared the preferences recorded on the
        checklist against those on the scanned image of the ballot paper and
        those in the related XML file;
    \item The IT security team compiled, investigated and reported on the
        findings.
\end{itemize}

The compliance inspection report outlined that a total of 1,510 ballot papers
were inspected and 4 processing errors were identified.  This seems to indicate
an error rate of less than 0.3\% per ballot.  Although it wasn't recorded how
many preferences were on each ballot, it seems to indicate a very small
per-digit error rate.  However, a careful reading of that experimental
description shows that the officials verified only the numbers from 1 to 6.
Errors in later preferences were ignored.  So this estimate may substantially
underestimate the overall rate of error.

To estimate the per-digit error rate implied by these data, we assumed that all
of the 1,510 ballot papers that were inspected had six preferences marked on
them, giving a total of 9,060 digits.  We also assumed that the 4 `processing
errors' were each a single-digit error.  This gave a per-digit error rate of
0.04\%, with a 95\% confidence interval of (0.01\%, 0.11\%).

In reality, some proportion of these ballot papers were likely to be informal
and have fewer than six preferences marked.  Adjusting the above assumptions
based on reported rates of informality by the AEC\footnote{For example,
\url{https://www.aec.gov.au/Voting/Informal_Voting/senate/}} had negligible
impact on these estimates.

\subsection{Informal experiment}

For the 2019 federal election, we conducted an informal experiment amongst 15
of our colleagues to get a rough estimate of the `end to end' accuracy of the
Senate vote digitisation process.  Each of our colleagues decided on their
Senate vote ahead of the election and made a private record of it for later
comparison.  On polling day, they each carefully completed their Senate ballot
paper in accordance with their planned vote.  After the election, it was
possible to compare these against the electronic file of ballots published by
the AEC.  Each of our colleagues searched for a vote that matched their own
vote either exactly or very closely.

All of our colleagues voted below the line in Victoria.  Due to the very large
number of possible ways to vote below the line, each of their votes was
extremely likely to be unique.  In addition, the electronic file from the AEC
also recorded the polling booth for each ballot.  These two facts together
allowed each of our colleagues to easily identify their own ballot paper in the
file and be confident that it was indeed their own.  This was true even if the
match were not exact, since the next `closest' matching ballot would typically
vary substantially from each person's private record.

Of our 15 colleagues, 12 found ballots in the file that exactly matched their
own records. This indicates perfectly accurate digitisation.  The remaining 3
found a mismatch: each of them had a single one of their preferences recorded
differently in the file than in their private record.  These mismatches  could
be due to an error in the AEC digitisation process or to a transcription error
on the part of our colleagues.  However, they do give us at least a rough
estimate of accuracy.

What per-digit error rate does this imply?
We use the following assumptions:
\begin{inparaenum}[a)]
    \item Each ballot had votes below the line;
    \item All boxes below the line were numbered;
    \item All of the reported errors were for a single digit.
\end{inparaenum}
These assumptions maximise the number of possible digits and minimise the
number of errors, and thus will give the lowest possible error rate estimate.
There were 82 candidates for Victoria.  This gives $9 + 73 \times 2 = 155$
digits per ballot, which is $155 \times 15 = 2,325$ digits in total.  Out of
these, we have 3 single-digit errors.  These give a per-digit error rate of
0.13\%, and a 95\% confidence interval of (0.03\%, 0.38\%).
The error rate here captures any errors either by a voter or by the
digitisation process, so it provides a rough upper bound on the latter's error
rate.

\subsection{What is the state of the art in digit recognition error rate?}

Accurately recognizing handwritten digits by computer is an important
consideration for many applications where data crosses from the physical world
into the digital. The MNIST (Modified National Institute of Standards and
Technology) database is a large database of handwritten digits that is commonly
used for training image processing systems.  The database consists of digits
written by high school students and American Census Bureau employees, and
normalised to be represented as grayscale images of size $28 \times 28$ pixels.
The current state of the art approach~\cite{mnist} to this dataset has an error
rate of 0.18\%.\footnote{There is unpublished work claiming 0.17\%.} Care must
be taken with this result, which is on a well studied and well curated data
set. While Australian ballot papers have boxes marked where each number should
be filled in, not all digits written in practice fall completely within the
box. Nevertheless, this gives an accurate lower bound on pure computer-based
digit recognition accuracy.   The AEC process involves human inspection which
means that it may be able to achieve better overall digit recognition accuracy.

The errors in digit recognition are not uniform: some digits are easier to
confuse, for example 1 and 7.  Most work on digit recognition does not publish
the cross-digit confusion rates.  \autoref{tab:confus} gives a confusion table
showing the percentage of each actual digit versus its predicted value from
experiments reported by Toghi \&
Grover~\cite{DBLP:journals/corr/abs-1809-06846}.  The overall digit recognition
error in this work is 0.89\%, which is substantially greater than the best
results reported above.

\begin{table}[t]
\caption{\textbf{Pairwise error digit rates.} The entry for row $x$ and column
    $y$ gives the percentage chance of (mis)recognizing a digit $y$ as a digit
    $x$. A dash `$-$' indicates less than 0.01\% chance of misrecognition.}
\label{tab:confus}

$$
\begin{array}{lc|rrrrrrrrrr}
& \multicolumn{10}{c}{\mbox{Actual}} \\
\multirow{11}{*}{\rotatebox{90}{Predicted}} & \mbox{Digit}
    &     0&     1&     2&     3&     4&     5&     6&     7&     8&     9\\
\hline
&0  & 99.22&     -&  0.08&  0.02&  0.10&  0.04&  0.14&     -&  0.06&  0.20\\
&1  &     -& 98.75&  0.14&     -&     -&     -&     -&  0.40&  0.04&  0.08\\
&2  &  0.12&  0.28& 99.56&  0.24&     -&     -&     -&  0.18&  0.02&  0.10\\
&3  &     -&     -&  0.22& 99.50&     -&     -&     -&  0.24&  0.14&  0.22\\
&4  &  0.16&  0.16&     -&     -& 98.65&  0.08&  0.10&     -&  0.12&  0.30\\
&5  &     -&  0.02&     -&     -&     -& 99.52&  0.22&  0.10&  0.18&  0.12\\
&6  &  0.10&  0.12&     -&     -&  0.06&  0.08& 99.48&     -&  0.14&     -\\
&7  &  0.08&  0.42&     -&  0.16&     -&  0.02&     -& 98.90&     -&  0.38\\
&8  &  0.10&  0.06&     -&     -&  0.48&     -&     -&     -& 99.16&  0.26\\
&9  &  0.22&  0.20&     -&  0.08&  0.72&  0.26&  0.06&  0.18&  0.14& 98.34\\
%&0  & 99.22&  0.00&  0.08&  0.02&  0.10&  0.04&  0.14&  0.00&  0.06&  0.20\\
%&1  &  0.00& 98.75&  0.14&  0.00&  0.00&  0.00&  0.00&  0.40&  0.04&  0.08\\
%&2  &  0.12&  0.28& 99.56&  0.24&  0.00&  0.00&  0.00&  0.18&  0.02&  0.10\\
%&3  &  0.00&  0.00&  0.22& 99.50&  0.00&  0.00&  0.00&  0.24&  0.14&  0.22\\
%&4  &  0.16&  0.16&  0.00&  0.00& 98.65&  0.08&  0.10&  0.00&  0.12&  0.30\\
%&5  &  0.00&  0.02&  0.00&  0.00&  0.00& 99.52&  0.22&  0.10&  0.18&  0.12\\
%&6  &  0.10&  0.12&  0.00&  0.00&  0.06&  0.08& 99.48&  0.00&  0.14&  0.00\\
%&7  &  0.08&  0.42&  0.00&  0.16&  0.00&  0.02&  0.00& 98.90&  0.00&  0.38\\
%&8  &  0.10&  0.06&  0.00&  0.00&  0.48&  0.00&  0.00&  0.00& 99.16&  0.26\\
%&9  &  0.22&  0.20&  0.00&  0.08&  0.72&  0.26&  0.06&  0.18&  0.14& 98.34\\
\end{array}
$$

\end{table}

\subsection{Analysing the election data (NOT simulations) to infer the error
rate}

We only have the reported ballots, not the ones that were ruled informal.
(Except of course we cannot distinguish human mistakes from scanning errors.)
Errors that make the vote informal are hidden.  

Recall that the formality rules require at least 6 unambiguous preferences
below the line, and that informal votes are not reported.  We can estimate the
number of hidden informal votes by observing the erroneous but formal ones.  We
use the number of repeated or missing numbers greater than 6 to approximate the
number of repeated or missing numbers less than or equal to 6.

\autoref{tab:missingBallots} shows the data, for BTL votes cast in Tasmania for
the 2016 Senate election.  The first column is the preference $p$ on the
ballot. The second  column is the number of ballot papers that contain $p$ more
than once.    The final column shows the number of ballots missing that
preference, showing preference $p-1$ and $p+1$ but not $p$. A 0 is not required
for $p=1$.  Note that there is a sudden drop at 12 because voters were
instructed to list at least 12 preferences, so many people listed exactly 12.
If the 12th preference was miswritten or misrecorded, then it did not count in
our table (there being no 13).

There would be no informal BTL ballots at all, and perfect zeros in the first 6
rows of \autoref{tab:missingBallots}, except for one special formality rule: if
there is \emph{also} a valid ATL vote present on the same ballot paper, then it
is counted instead, and both the valid ATL vote and the invalid BTL markings
are reported in the final database.  Hence we expect that the numbers in the
first 6 rows are only a small fraction of the ballots rendered informal by
either human or scanning errors.  There is a sudden increase at the 7th
preference, because  BTL votes with a repeated or omitted 7th preference are
still included in the tally, as long as their first 6 preferences are
unambiguous.

There are 97,685 published votes with BTL markings. Most of these were valid
BTL votes but some were only published because they had valid ATL votes as
well.  The most representative preferences are probably 7 to 9, being single
digits whose count is not artificially suppressed due to repetitions in them
causing the BTL vote to be informal and thus usually not published.  For these
preference numbers, the observed repetitions  are on the order of 0.5\%.  This
doesn't prove that the scanning process introduces errors at a rate of 0.5\%
per digit, because they could be caused by voter error.  It could also
underestimate the scanner error rate because it includes only those not
rendered informal. Nevertheless this provides an estimate of voter plus process
error.

\begin{table}[t]
\centering
\caption{\textbf{Counts of ballot papers with repeated and missed preferences.}
Tasmanian ballots with BTL marks, 2016.}
\label{tab:missingBallots}
\smallskip
\begin{tabular}{lrrrrrrrrrrrrrrr}
\toprule
\textbf{Preference} & 1 & 2 & 3 & 4 & 5 & 6 & 7 & 8& 9 & 10 & 11 & 12 & 13 \\
\midrule
\textbf{Ballots with preference repeated \phantom{1}}
& 573 & 385 & 303 & 231 &212 & 211 & 492 & 494 & 542 & 372 & 256 & 250 & 122 \\
\textbf{Ballots with preference skipped}
& 240 & 43 & 54 & 49 & 45 & 37 & 130 & 133 & 134 & 193 & 203 & 45 & 44 \\
\bottomrule
\end{tabular}

\end{table}

% -----------------------------------------------------------------------------

\section{Results}
\label{sec:results}

\subsection{Results from truncation and digit error models}

We simulated counts with errors using the ballot data for all 8 states and
territories from both the 2016 and 2019 Senate elections.  We used both the
truncation and digit error models, across a wide range of error probabilities.
For any given choice of model and error probability, we simulated 1,000
elections (each with their own random errors under that model).

For error rates between 0\% and 1\%, the only election for which we observed
any change in the elected candidates was for Tasmania in 2016.
%%%
This election was somewhat unusual in three ways. First, it was a very close
election, with the difference in tallies between the final two remaining
candidates, Nick McKim and Kate McCulloch, being only 141 votes. For
comparison, 285 votes were lost due to rounding. Second, there was a popular
labor candidate, Lisa Singh, who won a seat despite being placed fourth on the
party ticket, and the candidate above her not winning a seat. This means she
received many BTL votes specifically for her, rather than relying on ATL votes
for the party. Finally, the 2016 election was a double dissolution, which means
that twelve candidates were elected rather than the usual six.

\begin{figure}[t]
    \centering
    \includegraphics[width=\textwidth]{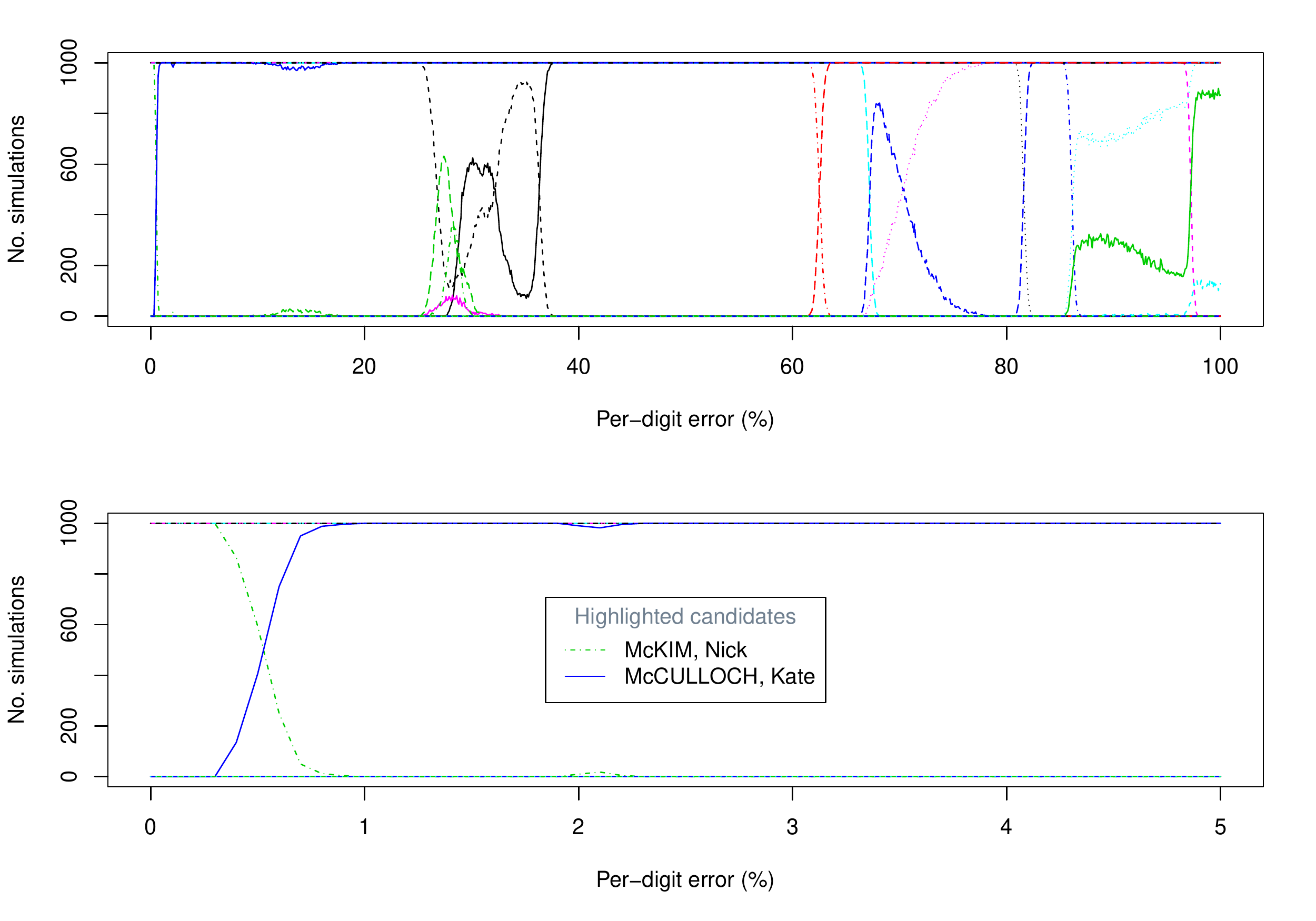}
    \caption{\textbf{Changing election outcomes as a function of error rate,
        Tasmanian Senate election 2016.}  The lower graph shows a complete
        reversal for a small error rate (about 0.5\%), between the state in
        which McKim wins consistently (no error) and that in which McCulloch
        wins consistently (1\% or greater error).  The upper graph shows
        similar behaviour for larger error rates---with error rates of more
        than 20\% there are sharp transitions between different election
        outcomes.}
    \label{fig:tas16}
\end{figure}

In the real election, the 12th (final) candidate that was elected was Nick
McKim.  In our simulations, once we introduced a small amount of error we saw
that a different candidate, Kate McCulloch, was sometimes elected instead.  As
we increased the per-digit error rate from 0\% to 1\%, we saw a complete shift
from one candidate to the other, see \autoref{fig:tas16}.  The truncation error
model led to the same outcome (data not shown).

\subsection{Pairwise digit error model}

We ran 1,000 simulations for Tasmania 2016 using the pairwise digit error
model.  Unlike the other models, we did not have a parameter to set but simply
used the pairwise error rate matrix shown in \autoref{tab:confus}.  This model
has an average per-digit error rate of 0.89\%.
%%%
Across the 1,000 simulations, we observed Kate McCulloch being elected 99.5\%
of the time, and Nick McKim for the remaining 0.5\%.  This is consistent with
the simple per-digit error model, which also resulted in Nick McKim
occasionally being elected when the per-digit error was comparable.

\subsection{Sharp transitions}

The fact that such a sharp transition happens from electing one candidate to
another was initially surprising to us.  Rather than simply `adding noise' and
leading to randomness in which candidates got elected, the noise seems to be
leading to a systematic bias in favour of or against specific candidates.  This
behaviour can be seen more clearly as the error rate is increased to larger
values (beyond values that would be plausible in practice), see
\autoref{fig:tas16}, where sharp transitions are visible also at 28\%, 36\%,
62\%, 68\%, 82\%, 86\% and 97\%. 

To investigate possible reasons for this, we looked at how individual ballots
were affected by the simulated errors.  Compared to the no-error scenario, two
broad types of outcome are possible:
\begin{itemize}
\item The ballot becomes informal and is not counted.  This will happen when it
    does not meet the formality requirements, e.g., does not have at least a
    single first preference above the line or consecutive preferences numbered
    1 to 6 below the line.
\item The ballot ends up exhausting before reaching a candidate.  This will
    happen if the preference order becomes disrupted due to an error, which has
    the effect of truncating the preferences and not enabling the ballot to be
    counted in favour of any candidates further down the preference list.
\end{itemize}
We investigated these effects in the context of the Tasmanian 2016 election;
we report on this in the next few sections.  We found that the first type of
effect was the dominant factor in determining the election outcome.

\subsection{Why random errors affect different candidates differently (Tasmania
2016)}

We saw earlier that for small error rates, we have either Nick McKim (from the
Australian Greens party) or Kate McCulloch (from the One Nation Party) elected
as the final candidate.
%%%%
There were 339,159 formal ballots for this election.  For each one, we looked
at the preferences to see:
\begin{itemize}
\item whether it was an ATL or a BTL vote,
\item which of the above two candidates (or their respective parties, if it was
    an ATL vote) was more highly preferred, or neither one.
\end{itemize}
%%%
\autoref{tab:tas2016} shows how the ballots split into these categories.  The
most important fact to note is the relative number of ATL and BTL votes in
favour of each candidate: more than 80\% of the ballots in favour of McCulloch
were ATL votes, while for McKim it was less than 70\%.

\begin{table}[t]
\caption{\textbf{Partition of the Tasmanian 2016 ballots.}  The number of
ballots split by whether it is an above-the-line (ATL) or below-the-line (BTL)
vote, and which candidate (if any) out of Kate McCulloch or Nick McKim is
preferred over the other.}
\label{tab:tas2016}
\smallskip
\centering
\begin{tabular}{lrrr}
\toprule
& \textbf{McCulloch} & \textbf{McKim} & \textbf{Neither}  \\
\midrule
\textbf{ATL} & 73,975 & 97,331 & 72,468  \\
\textbf{BTL} & 17,066 & 42,170 & 36,149  \\
\bottomrule
\end{tabular}
\end{table}

\begin{figure}[t]
    \centering
    \includegraphics[width=0.5\textwidth]{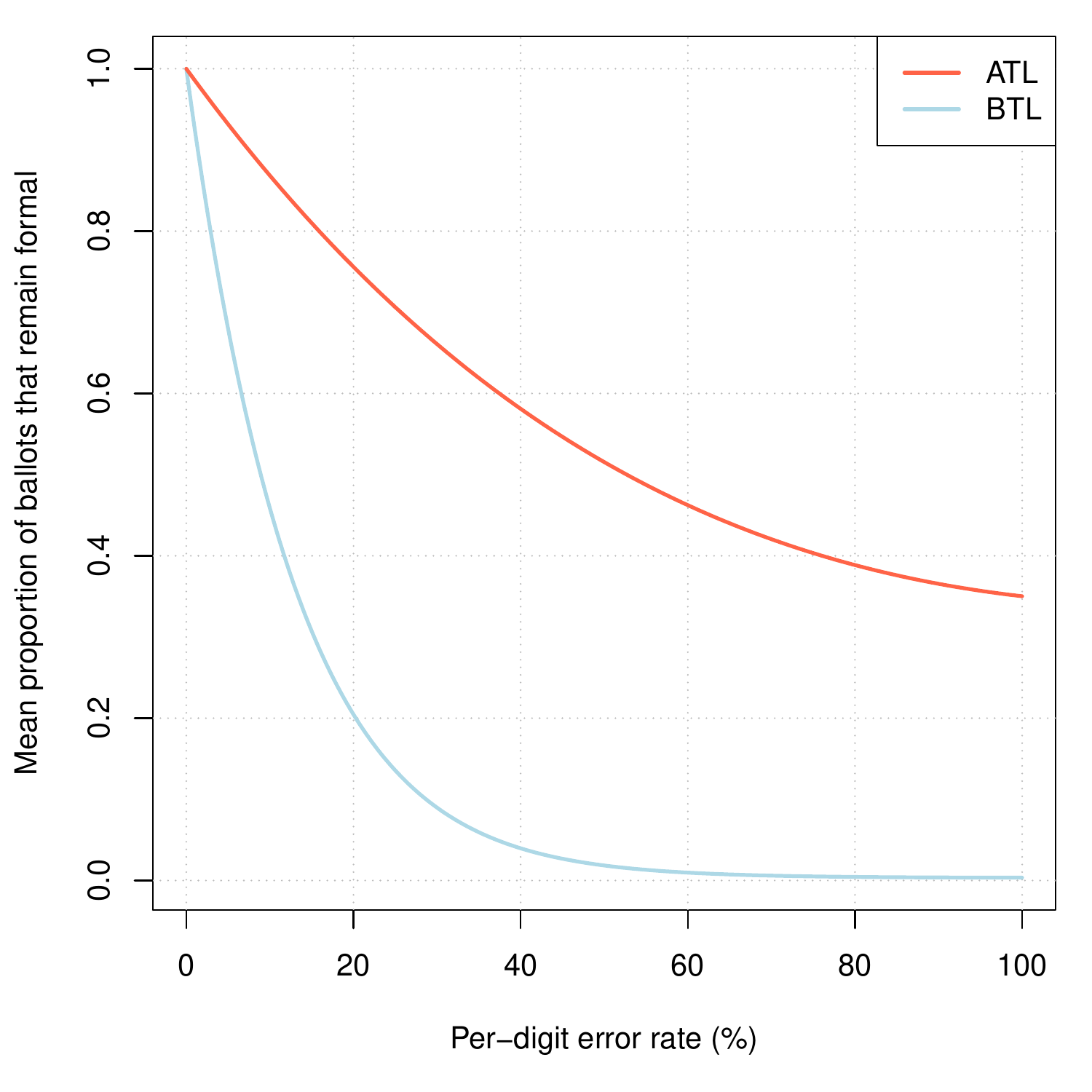}
    \caption{\textbf{Effect of the per-digit error rate on the formality of
    votes.} The impact on above-the-line (ATL) and below-the-line (BTL) votes
    are shown separately.}
    \label{fig:tas16-mean-prop-formal}
\end{figure}

When errors are introduced, ballots that were BTL votes were much more likely
to become informal.  \autoref{fig:tas16-mean-prop-formal} illustrates this: the
larger the error rate, the greater the disparity in how many of the ATL or BTL
ballots became informal.  This on its own is enough to explain the systematic
shift from McKim to McCulloch as error rates increase.

\begin{figure}
    \centering
    \includegraphics[width=\textwidth]{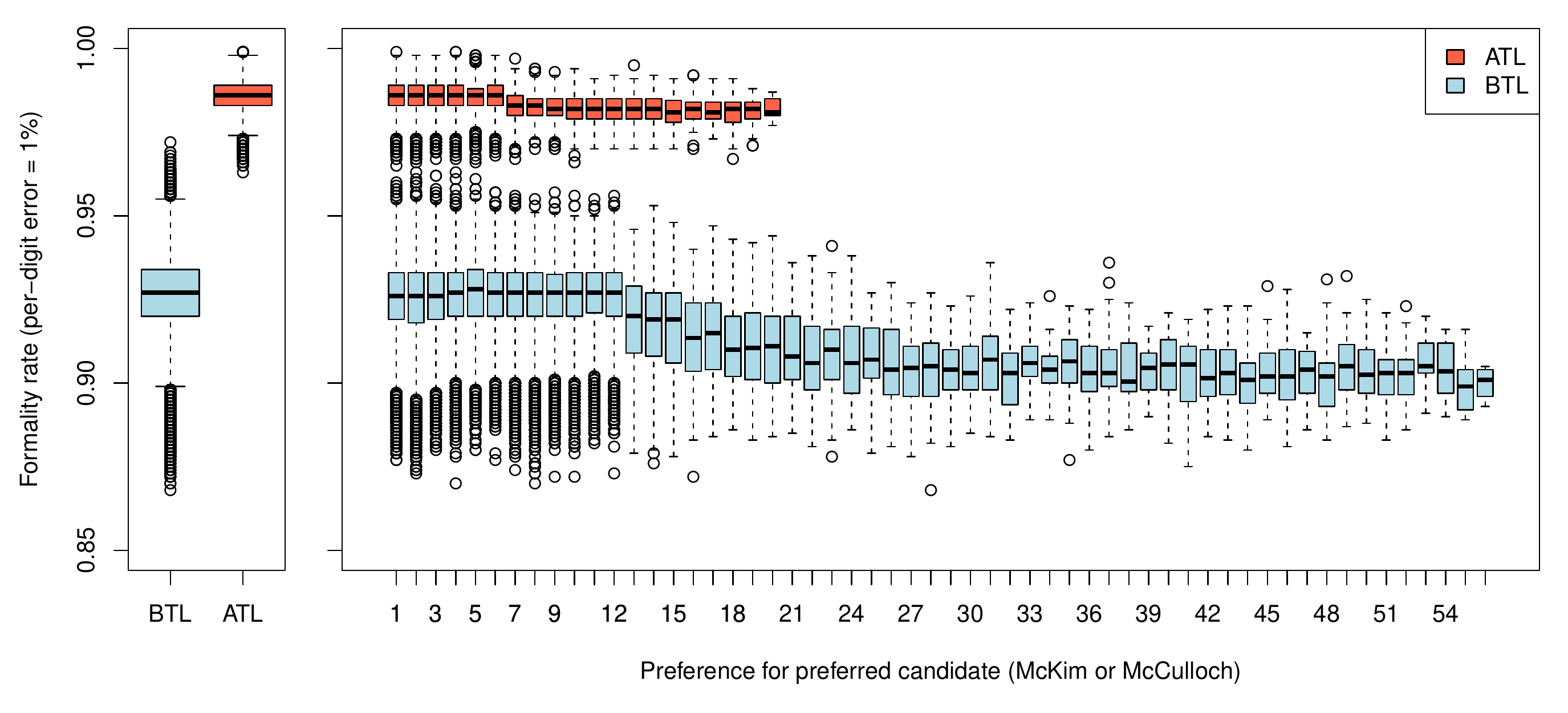}
    \caption{\textbf{Formality rates for votes with random errors injected.}
    These are split by ATL/BTL (left panel) or by the position of the preferred
    candidate (right panel).  Error rates vary greatly between ATL and BTL
    votes, but not much between preferences within those categories.}
    \label{fig:tas16-formality}
\end{figure}

For more insight, we took a closer look at the simulations that used a
per-digit error rate of 1\%.  For each ballot, we define the \emph{formality
rate} to be the proportion of simulations for which it remained formal.
\autoref{fig:tas16-formality} shows the distribution of the formality rate
across different types of ballots.  The left panel shows the clear disparity
between ATL and BTL votes.  This reiterates the difference we saw on average
from \autoref{fig:tas16-mean-prop-formal}, but in addition we see that this
disparity is very consistent across individual ballots (from the very little
overlap for the ATL and BTL ballots).

When we further divided the ballots based on where in the preference list the
voters placed their preferred candidate out of McKim or McCulloch, the
distribution of formality rates was relatively consistent (right panel of
\autoref{fig:tas16-formality}).
%%%
This indicates that the major factor leading to McCulloch replacing McKim is
simply the lower formality rate for BTL votes, after random errors were added,
coupled with the fact that a larger proportion of ballots in favour of McKim
were BTL votes.

For the less plausible larger errors, the sharp transitions came from new
effects causing biases against major parties, who lost out as randomisation of
preferences reduced their typical large first preference collection. This also
caused major parties to not get multiple candidates elected in the first
counting round, which meant that major party candidates low down on the  party
ticket tended to get eliminated before they could get preferences passed on to
them, as they were reliant on BTL votes to avoid being eliminated before the
first candidates of minor parties who could get ATL votes.

\subsection{Varying the formality requirements}

The formality requirements differ for ATL and BTL votes.  In particular, BTL
votes require at least 6 consecutive preferences in order to be declared
formal, whereas ATL votes only require a single preference.  This is one reason
why the formality rate for BTL is lower once errors are introduced.

We investigated whether changing the formality rules could ameliorate the
systematic bias caused by the introduction of errors.  Specifically, we varied
the number of consecutive preferences required for a formal BTL vote, ranging
from 1 (i.e.\ the same as ATL votes) to 9 (i.e.\ more stringent than the current
rules).

\begin{figure}[t]
    \centering
    \includegraphics[width=\textwidth]{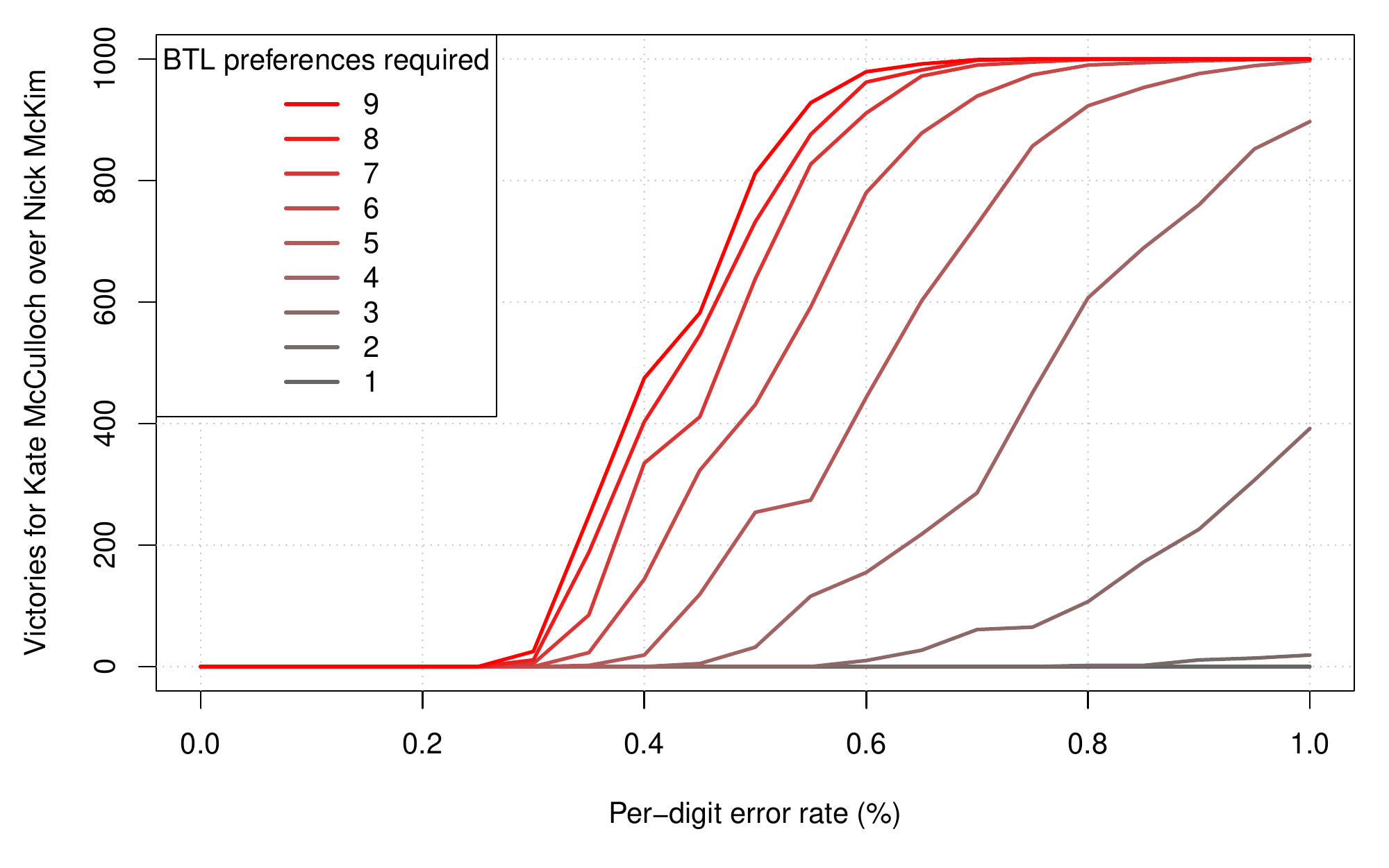}
    \caption{\textbf{The effect of formality rules on election outcomes.}  As
    the number of preferences required for a valid BTL vote increases, so does
    the rate at which BTL votes are excluded due to random errors.  This
    produces a faster transition from one winning candidate to another as the
    error rate increases.}
    \label{fig:formality-experiment}
\end{figure}

\autoref{fig:formality-experiment} shows the impact of these choices on  how
often McCulloch was elected instead of McKim.  Making the formality requirement
less stringent reduced the bias, and once the formality rules were aligned for
ATL and BTL votes, the election result remained mostly unchanged even in the
presence of errors.

\subsection{Truncation of preferences}

Other than causing ballots to become informal, errors can result in votes not
being counted for certain candidates if the error truncates the preference
order.  Candidates who obtain more of their votes from later (higher-numbered)
preferences should be more  affected by such truncation.

We investigated whether this might be occurring in our simulations.  For each
ballot, we compared the number of valid preferences before and after simulated
errors.  There was a clear signal of truncation: ballots that had around 60
valid preferences (which were all BTL) only had on average around 30 valid
preferences remaining when the per-digit error was set to 1\%.  In contrast,
ballots that had 10 valid preferences (irrespective of whether they were ATL or
BTL) maintained  almost 10 valid preferences on average.

While this extent of truncation is stark, it might not necessarily lead to any
change in the election outcome because many of the later preferences might not
actually be used during the election count.

\begin{figure}[t]
    \centering
    \includegraphics[width=\textwidth]{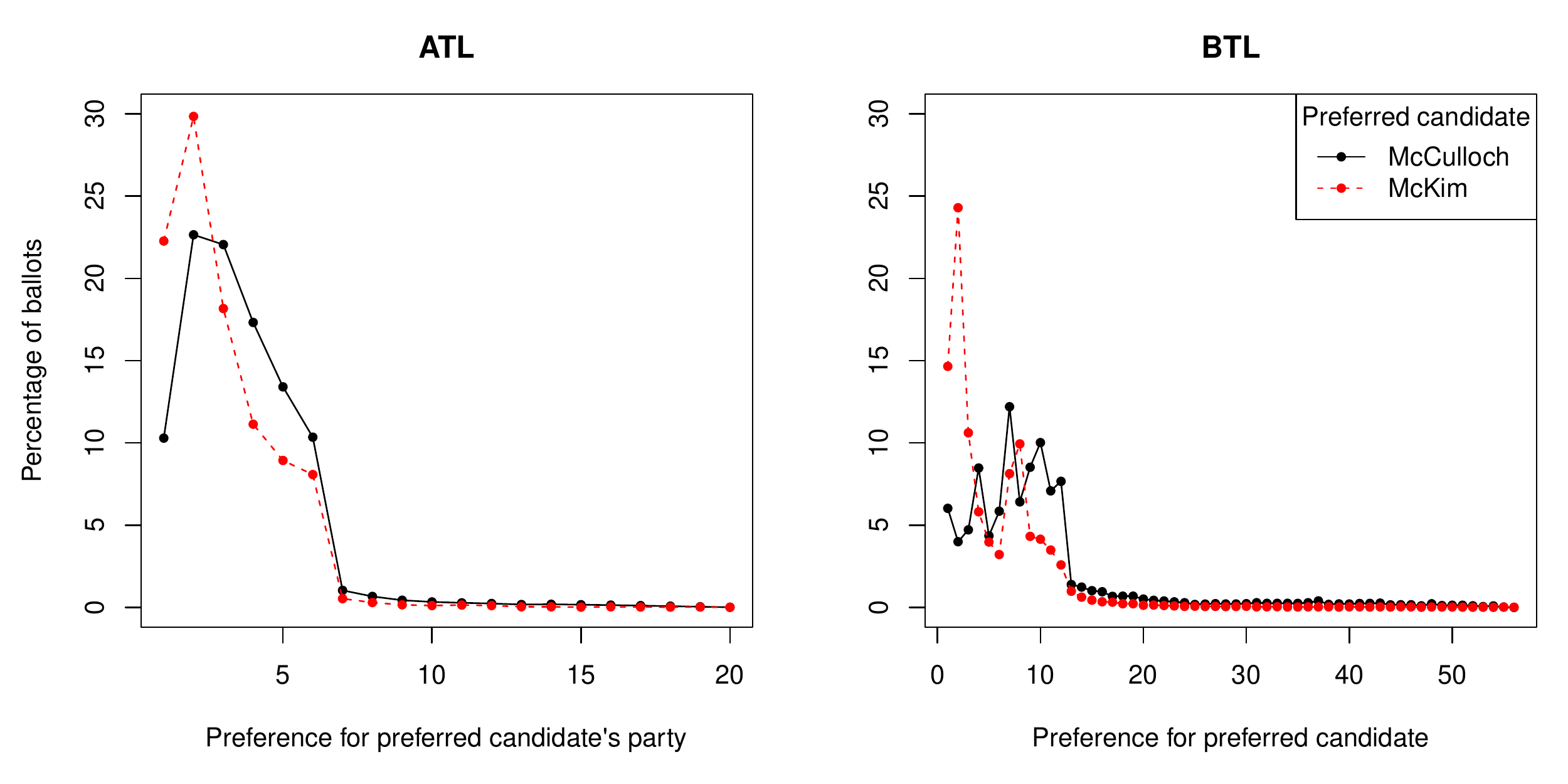}
    \caption{\textbf{Histograms of preference number.} These are shown for a
    candidate or party depending on whether the votes are above or below the
    line.}
    \label{fig:prefdist}
\end{figure}

In the case of the Tasmanian 2016 election, we looked at ballots in favour of
each of McKim and McCulloch to see whether they tended to get their votes from
earlier or later preferences.  \autoref{fig:prefdist} shows the distribution of
these.
%%%
Interestingly, we see that McCulloch relies more on later preferences than
McKim.  Therefore, it is McKim rather than McCulloch that should benefit from
any truncation effect.  This works in the reverse direction of the
formality-induced bias described earlier, however the truncation did not act
strongly enough to reverse that bias.

% -----------------------------------------------------------------------------

\section{Concluding remarks}
\label{sec:conc}

We are not aware of any previous study of the the effects of random errors in
digitization on election outcomes.  While there is a considerable body of work
on margin of error for polling, there is little study of the effect of errors
on elections.  Richey~\cite{richey} examines how `errors' in voting can effect
elections, but here the error is that a voter votes for a party that does not
represent their best interests.
\ignore{In that sense the work is completely unrelated.  The conclusion is that
campaign spending by Republicans is able to "confuse" voters into voting
Republican when that is not in their best interest, and this can change the
result of elections.}

The previous section clearly demonstrates that random errors during counting do
not necessarily lead to `random' changes to election outcomes.  We were very
surprised by the sharp transitions in election results as error rates changed,
illustrated in \autoref{fig:tas16}.  Systematic biases can arise due to
interactions with the election rules.

For Australian Senate elections, a key factor is the formality requirements.
BTL votes have more stringent requirements, which ends up creating a systematic
bias against BTL votes in the presence of random errors. Candidates who rely on
BTL votes (e.g.\ if they are relying on their individual popularity) will be
more affected by random errors than those relying on ATL votes (e.g.\ via
membership of their party).  Changing the formality requirements to reduce the
disparity between ATL and BTL votes also reduces this bias.

Candidates who rely on accumulating later preferences are more affected by
random errors than candidates who rely primarily on their first-preference
votes.  However, this effect was much weaker than the bias induced by
differences in formality requirements.

These results raise questions about how formality rules should be specified in
order to be fair to candidates with different voting patterns.  More relaxed
formality rules could be applied which are less likely to have strong
differences across different kinds of votes. For example, a BTL vote could be
formal if the first 6 most preferred candidates are clear, even if they are not
numbered from 1 to 6, e.g.\ a vote with preferences 1, 2, 4, 5, 6, 7 and no
preference 3 still gives a clear ranking of the first 6 candidates.

In this paper we  consider only Australian Senate elections with their
particular ATL/BTL voting mechanism.   Two lessons can be taken from this
exercise to other forms of voting.
%%%
First, if there are two or more forms of ballot and the rules for formality are
different for these different forms of ballot, then random errors may affect
the different forms differently, regardless of whether the voter can choose
their form or different voters are assigned to different forms. This is
applicable to any kind of election whether plurality voting or ranked voting.
%%%
Second, considering elections where voters rank candidates with only one form
of ballot, e.g. standard STV, Borda, or Condorcet elections, assuming the rules
of formality are such that the ballot is truncated when the ranking becomes
uninterpretable, then candidates relying on accumulating later preferences will
be more affected by random errors than other candidates.  But we do not have a
real world case that illustrates that truncation errors alone lead to a change
in a result.

% -----------------------------------------------------------------------------

\subsubsection*{Acknowledgements.}

We would like to thank our colleagues who participated in our informal
experiment during the 2019 Australian federal election.  Thanks also to Philip
Stark for very valuable suggestions on improving the paper.

% -----------------------------------------------------------------------------

\bibliographystyle{splncs04}
\bibliography{BIB}

\end{document}